# Why IP-based Subject Access Requests Are Denied?

Supriya Adhatarao
Univ. Grenoble Alpes, Inria, France
supriya.adhatarao@inria.fr

Cédric Lauradoux
Univ. Grenoble Alpes, Inria, France
cedric.lauradoux@inria.fr

Cristiana Santos
Utrecht University, Netherlands
c.teixeirasantos@uu.nl

## ABSTRACT

Understanding the legal status of IP addresses is complex. In Europe, the General Data Protection Regulation (GDPR) is supposed to have leveraged the legal status of IP addresses as personal data, but recent decisions from the European Court of Justice undermine this view. In the hope of providing more certainty, we have looked on *how 109 websites deal with IP addresses*. First, we analyzed the privacy policies of these websites to determine how they considered IP addresses. Most of them acknowledge in their privacy policy the fact that IP addresses are personal data. Second, we submitted subject access requests based on the IP addresses used to visit different websites. Our requests were often denied. Websites justify their answers with different explanations suchlike *you need to register*, or *IP addresses do not allow to identify you*, to name a few. If IP-based SARs are often denied, it creates an opening door for websites to track their accountless users without being accountable because users cannot exercise their rights (access, erasure etc.). We make several proposals to improve this situation by modifying how IP addresses are allocated to a user.

## 1 INTRODUCTION

An IP address is an online identifier [4, 26] used for identifying devices online and to route information between websites and its end-user devices. IP addresses can be effectively used to track Internet users [37], as cookies [22, 30] or browser fingerprinting [32]. Almost every website collects and processes IP addresses of their users and yet the question remains whether *they are considered as personal data by law?* The answer to this question has surprisingly deep consequences for online privacy and on the way websites operate. *If the answer is yes*, then IP addresses should only be "collected for specified, explicit and legitimate purposes and not processed for purposes incompatible with the purposes for which they were originally collected" (Article 5(1)(a)(b) GDPR). The processing of IP addresses would then have to be "adequate, relevant and not excessive in relation to the purposes for which they are collected and/or further processed" (Article 5(1)(c) GDPR).

*If the answer is no*, the burden relies on the side of Internet users. Therefore, websites prefer to consider that IP addresses are not personal data.

Empirical results and the current legal reasoning on IP addresses are not convergent. Notably, empirical studies have demonstrated [33, 37] that users can get assigned over time a set of IP addresses which are unique and stable. According to these finding, if IP addresses are not personal data, then users face the risk to be tracked and potentially identified by websites without adequate protection. From the legal community, there is a long-term trend wherein personal data has been constantly broadened to include IP address [6, 7, 15, 28]. Court decisions can be summarized as follows: *IP addresses are personal data under certain conditions.* At first sight, this position looks like a perfect compromise for Internet users and websites: users are comforted with the fact that it is not possible to use IP addresses to track them, and websites can find exceptions to keep collecting IP addresses without being bound to the legal requirements set forth by the GDPR.

In this paper, we try to find whether this compromise has favored the websites or the rights of Internet users. We attempted to determine whether IP addresses are personal data using the tools that are available to an Internet user – *exercising a subject access request*. We visited 109 websites of public organizations and private companies. We analyzed their privacy policies and then sent subject access requests. Our requests were very specific: *we asked websites to provide all the data concerning the IP addresses used while visiting their websites.*

The results obtained are clear: companies consider IP addresses as personal data, though it seems there is no implications in practice. Some websites were not consistent: they acknowledge in their privacy policy the collection and processing of IP addresses, but claim having nothing about the requester. Some websites also consider that they are not collecting personal data if the user is not logged in. The most interesting answer stated that *a given IP address can identify several users*. This answer is legitimate because so far Internet users are unable to prove that they have used a given IP address during a certain period of time and therefore, they cannot submit a valid subject access request to websites.

From the obtained answers it can be reasoned that the current status of IP addresses has favored websites in detriment of Internet user's rights. Conversely, the GDPR aims to provide transparency to the processing of personal data and, in particular, online tracking. As users visit websites both when they are registered and non-registered, in the recurrent occasion of not being registered, the IP address is the most obvious identifier that is known by both users and websites. Users expect to request websites to provide all the data related to his/her IP address. Hence, IP-based subject access requests is the most feasible tool to understand the extent of online tracking compliance. As such, if users cannot submit IP-based subject access request (SAR), they are hence disempowered to access their own personal data, cannot verify the lawfulness of the processing and tracking, nor are they able to fulfil their rights (deletion, rectification, restrictions, etc.).

To fix this situation, one solution would be to change how IP addresses are allocated. It will require to allocate IP addresses to people instead of electronic devices. But this is not realistic because it is not compatible with the current Internet mechanisms and it will be difficult to make it scale. Instead, we propose that Internet users can request *certificate of usage* for their IP addresses. This certificate could be obtained from Internet Service Providers with a verification by national data protection authorities.

The paper is organized as follows. We explain the motivation for IP-based subject access request in Section 2 and current legal positioning concerning IP addresses in Section 3. Section 4 describes the websites visited during our analysis and the methodology used. Section 5 explains what websites say about IP addresses in their privacy policies. Section 6 details how websites have denied our IP-based subject access requests. Section 7 suggests solutions to fix IP-based subject access requests.

## 2 MOTIVATION

Users studies [34, 47] have shown that Internet users consider that IP addresses are very sensitive information that need to be protected. Indeed, IP addresses of user devices can already provide many information on the user like his/her geographical location [33, 42]. Concerns of Internet users are legitimate because IP addresses are used by websites to track their online activities as shown in [22, 43, 49].

### 2.1 Why IP-based SARs are important?

Internet users can be divided into two categories: registered users and casual/accountless users. These categories are not exclusive since a person can assume both roles at different times for a website. However, these two kinds of users are very different form a website's perspective. It is easier for the website to interact with a registered user because he/she has setup an account with a login and a password that establishes his/her identity on that website. Many websites are nowadays allowing their registered users to access all their personal data using privacy dashboards like Google TakeOut[1]. These dashboards are an implementation of the right to access defined in the GDPR and in many other regulations.

To illustrate this scenario, we setup an account on Google on a computer and started to visit Google's websites while being a registered and also as a non-registered user. Then, we downloaded the data using Google TakeOut with our account. It appears that the downloaded data describe the different collection and processing of personal data. It includes the IP addresses used to visit Google's websites along with other information for each access made as a registered user. However, no information was found about the visits made as an accountless user. It means that Google has not associated the browsing activities performed using our computer while being unregistered to our account, and as such, this method of access does not include data while being unregistered.

Our experiment with Google TakeOut leads us to the following question: *Is there a method to access the data collected by website (e.g. Google) for an accountless user?* This question is central in our work. An accountless user needs to provide identifiable elements to help websites (or any data controller) to identify his/her personal data. These elements can consist on some identifiers and the obvious choice in our case is the user's IP address because this address is shared by both the accountless user and the website (*e.g.* Google). This is important to understand if accountless users are tracked or not because accountless users have also a right to transparency. They have the right to know how their data are being used by websites.

### 2.2 IP address allocation does not help

When a user visits a website on her browser, packets are exchanged between the user device and the server hosting the website. Source and destination of those packets are identified by IP addresses. These IP addresses can be either IPv4 or IPv6, it is irrelevant for the outcomes of our research. However, it is relevant to understand how IP addresses are allocated to user devices and which IP addresses are seen by the websites. This helps to understand if a data controller should provide data w.r.t to an IP address and the way IP-based subject access request needs to be managed.

It is important to understand that IP addresses are allocated to user's devices and not directly to a given user. This is important because if a device is shared by multiple users, then it is not possible to establish a link between an IP address and a given user. Network topologies and IP addresses allocation schemes [42] also create situations in which several users devices are going to be attributed the same IP addresses or at least websites are going to observe the same IP addresses when they are visited. For instance, an IP address can be allocated to a given user's device by a WiFi router at time $t$ and it can be allocated to a different device at time $t + 1$.

---

[1] https://takeout.google.com

In recent years there is a growing concern towards online privacy. Virtual private network (VPNs) and Tor[2] are privacy enhancing technologies which can be used to hide IP addresses used by people when visiting websites. This supports also the idea that IP addresses are sensitive information for Internet users. When users submit a IP-based SAR using VPN/Tor IP addresses, it is impossible for the data controller to identify a user as many people share the same IP addresses using VPN and Tor.

## 3 ARE IP ADDRESSES PERSONAL DATA?

Personal data is *"any information relating to an identified or identifiable natural person* (Article 4(11) GDPR). "Identified" means when a person, within a group of persons, is "distinguished" from all other members of the group [18]. "Identifiable" person is one who, although has not been identified yet, is possible to be identified in the future, either directly or indirectly. Such possibility to be identified can be made in two ways: i) by reference to identifiers: as a name, an identification number, location data, an online identifier or to one or more factors specific to the physical, physiological, genetic, mental, economic, cultural or social identity of that natural person (Article 4(11) GDPR [23]); and ii) by *"all the means likely reasonably to be used either by the controller or by any other person to identify the said person"* (Recital 26 GDPR).

*Does this definition of personal data apply to IP addresses?* The answer to this question is very important and has been debated for years by both computer scientists and lawyers. We explain the current state of the debate.

Let us consider a case study wherein Alice has subscribed to an Internet Service Provider (ISP) called Bob. *Is Alice's IP address personal data for Bob?* There is a consensus [5, 14] acknowledging that Alice's IP address is personal data for Bob. Indeed, Bob stored the identification data of all his subscribers (including Alice's) and assigned them their IP addresses. Therefore, Bob can identify Alice from her IP address.

Now, Alice visits Eve's website. *Is Alice's IP address personal data for Eve?* The answer to this question divides the data protection scholarship [45, 50]. This community diverges in the understanding of "dynamic" or "temporary" IP addresses as personal data[3] [3, 21, 24, 25, 31, 36, 46, 50]. In 2011, an official report of the Publications Office of the EU [8] studied the case law on the circumstances in which IP addresses are considered personal data; it showed that from 49 decisions regarding the IP address's status in 13 EU Member States, 41 decisions ruled (either explicitly or implicitly) that IP addresses should be considered personal data and 8 ruled against this interpretation.

Currently, three stances prevail, as summarized in Table 1, which presents a non-exhaustive list of the legal positions from courts and stakeholders (EDPS and 29WP) on the legal status of IP addresses.

(1) IP addresses can *per se* identify a person;
(2) IP addresses do not suffice alone, and thus additional information are needed to enable the identification of a person;
(3) IP addresses only configure personal data if additional data is obtained by lawful means.

### 3.1 IP addresses *alone*

In a case held at the European Court of Human Rights (ECHR) [17], the decision of the court evidences that it regarded dynamic IP address as information based on which the offender at stake could be identified. In Germany, both the Berlin district court and an appellate court decided that IP addresses are personal data; they added that "determinability" of a person should account for both legal and illegal means to obtain additional data [8]. The European Data Protection Board [1] declared that IP addresses should be treated as personal data by both ISPs and search engines (even if they are not always personal data) and adds that *unless an ISP or a search engine are in a position to distinguish with absolute certainty that the data correspond to users who cannot be identified, they will have to treat all IP information as personal data, to be on the safer side*. The European Data Protection Supervisor (EDPS) referred that for IP addresses to count as personal data, there is no requirement that the data controller knows the surname, first name, birth date, address (among other) of the individual whose activity it was monitoring. It further stated that an IP address which is showing special behaviour in terms of the transactions one can follow, then in a reasonable world, that is an individual [27].

### 3.2 IP address with additional information

The Court of Justice of the European Union (CJEU) determined in Breyer case [15] that dynamic IP addresses (temporarily assigned to a device), per se, are not information relating to an "identified" person, due to the fact that *"such an address does not directly reveal the identity of the person who owns the computer from which a website was accessed, or that of another person who might use that same computer"*. IP addresses can constitute personal data, provided that the relevant person's identity can be deduced from a combination of the IP address and additional identifiable data. This additional information [15, 19] can consist of *e.g.* name, login details, email address, username (if different from the email address), subscription to a newsletter, or other account data, in the course of logging in and using the website; cookies containing a unique identifier [20], device fingerprinting, or similar unique identifiers. By holding additional data, the website can tie it with the visitor's IP address, and therefore this visitor would be identifiable[15]. This argument explains why everyone would agree that Alice's IP address is a personal data for Bob (as ISP), because he knows both Alice's name. However, if Eve has access to additional identifiable information (uniquely identifiable), then Alice's IP address is personal data for Eve.

---
[2]https://www.torproject.org/

[3]Such divergence does not happen in the case of "static" or "fixed" IP, which are 'invariable and allow continuous identification of the device connected to the network' [15] (parag 36).

| Decisions | IP address are personal data | | |
|---|---|---|---|
| | Alone | With added info. | Added info obtained by legal means |
| EDPS | ● | - | - |
| Berlin district court | ● | - | - |
| 29 WP | ● | - | - |
| CJEU (Breyer) | - | ● | ● |
| ECHR | ● | - | - |
| Munich court | - | - | ● |
| Paris appeal court | - | - | ● |

Table 1: Summary of the legal positions concerning the status of IP addresses as personal data.

### 3.3 IP address with lawfully obtained additional information
.

Pursuant to this view, an IP address will only be personal data when a website has legal means to lawfully obtain access to sufficient additional data held by a third-party in order to identify a person. Respectively, IP addresses will not consist of personal data when such added data is obtained in a way that is prohibited by law, because ISP have to meet its own legal obligations before it just hands over the data. As ISPs are generally *prohibited* from disclosing information about a customer to a third party, the only means wherein an ISP is forced to disclose IP addresses data consist of consent, court order, by law enforcement agencies or national security authorities [19]. The Paris appeal court, in two rulings, stated that the processing of IP addresses does not constitute personal data unless a law enforcement authority obtains a user identity from an ISP [6, 7]. The Munich district court, in 2008, held that dynamic IP address lack the necessary quality of "determinability" to be personal data, which means that it cannot be easily used to determine a person's identity, without a significant effort and by using "normally available knowledge and tools." The court recalls that ISPs are not legally permitted to hand over the information identifying an individual, without a proper legal basis (only when ordered by a court) [38].

The CJEU in Breyer case [15] concluded that a dynamic IP address constitutes personal data if the website operator has "legal means" for obtaining access to additional information held by the ISP that enables the website publisher to identify that visitor, and there is another party (such as an ISP or a competent authority) that can link the dynamic IP address to the identity of an individual. *Legal means* could consist, for example, bringing criminal proceedings in the event of denial-of-service attacks to obtain identifying information from the ISP.

### 3.4 Empirical studies
Against this background, empirical studies have already demonstrated in [33, 37] that a user can get assigned, over time, a set of IP addresses which are unique and stable. Mishra *et al.* [37] found that the retention period of an IP address was, on average, 9.3 days. 2% of user's IP addresses did not change for more than 100 days and 70% of users (amounting to 1569) had at least one IP address constant for more than 2 months. Therefore, it is possible to discriminate Internet users based on their sets of IP addresses. Cycles and patterns of IP addresses were also observed in [37] in a user's browsing history. These cycles have the potential to be abused to infer traits of user behaviour, as well as mobility traces.

**Summary** Court decisions (both at national and CJEU level) and stakeholder positions so far diverge on the legal status of IP addresses. Conversely, empirical studies make evident that even dynamic IP addresses are afforded with uniqueness and stability features and thus could be a relatively reliable and robust way to identify a user visiting a website.

This ambiguity triggers uncertainty and confusion to all the organizations handling IP addresses.

## 4 TARGETED WEBSITES

Our research focuses on inquiring whether websites consider IP addresses as personal data or not. We have chosen 124 organizations and performed experiments to check how they handle IP addresses. We have visited two groups of websites: 74 websites maintained by private companies and 50 websites maintained by public organizations from which most are national Data Protection Authorities. We analyzed the privacy policies of these websites to check if they mention the processing of IP addresses. Then, we have submitted subject access requests based on the IP addresses used to visit the corresponding websites.

### 4.1 Private companies websites

Firstly, we have chosen 22 websites of private companies that are considered the most visited around the world[4] in 2021. Article 29 Working Party (29WP) stated that *devices with a unique identifier (through the cookie) allows the tracking of users of a specific computer even when dynamic IP addresses are used. In other words, such devices enable data subjects to be 'singled out', even if their real names are not known.* Hence, our next choice was a list of 52 websites of companies that set cookies on their user's browser. The computation of these cookies depends on the IP address of the user. We were able to figure this out using a guess and determine approach. First, a website is visited and a cookie is set. Then, the browser is reset and the cookie is removed from the browser but kept in our log. Then, one parameter in the computer's setup is changed and the website is visited again. A new cookie is obtained and compared with the previous one from our log. Complete reverse engineering seems difficult. This methodology has been defined by [22].

Table 6 in Appendix 9.2 provides the names of all the 74 private companies we have considered in our work.

### 4.2 Public organizations websites

We have chosen 50 websites from Data Protection Authorities (DPAs), the website of the European Data Protection Board (EDPB) and the website of the European Data Protection Supervisor (EDPS). DPAs are independent public organizations that supervise through their investigative and corrective powers, the application of the data protection law in each EU country. They all have a website, therefore it was logical to investigate how do they consider IP addresses. We have considered all the DPAs listed by the GDPR hub[5].

We have also visited the website of the EDPB [6] and the EDPS [7]. Table 7 in Appendix 9.3 provides a list of 48 DPAs whose websites we visited during our experiments.

### 4.3 Visit's details

We have visited the websites with three different IP addresses. We have used the default dynamic IP address provided by our ISP and we have also requested a static IP address to the same ISP for our device. We visited also the websites through Tor Network[8]. For the websites, using Tor means that we are using the IP address of the Tor exit node. It is likely that many devices and users use the same exit node and therefore the same IP address. Even though we have used different IP addresses to access the websites, due to the fact that our request was always denied, we did not mention the use of different IP addresses in our further discussions.

All the visited websites could be viewed as an external user, but some of them could be also accessed as a registered user. We have always visited the websites as an *external user*. In fact, we have created an account on 20 private websites (Google, Youtube, Amazon, LinkedIn, Reddit, Zoom, Yahoo, Ebay, Pinterest, Wikipedia, Twitter, Twitch, Roblox, Bitly, Fandom, Tripadvisor, Microsoft, Apple, Facebook and Indeed) to observe if it would have changed the response obtained to our subject access requests. However, even with a registered user account, our requests were *denied*. Hence, we only address the visit of websites as an external user in our further discussions.

### 4.4 Expectations concerning our requests

We did not expect any organizations to answer positively to our requests. We expected that data controllers would either reply that several devices can be associated to a given IP address or that several users could be associated to the device associated to the given IP address. We believe that it is not possible for a website to accept a subject access request based on an IP address, because it is very hard to verify that the IP address has been actually used by the data subject sending the request. If a website provides data to such a request, then there is a data breach. Security considerations mandate denial of any IP-based subject access requests. The fact that we were not optimistic about the obtained answers does not mean that we do not believe in IP-based subject access request. It means instead that we need new tools and protocols to submit them.

---

[4]https://ahrefs.com/blog/most-visited-websites/
[5]https://gdprhub.eu/index.php?title=Category:DPA

[6]https://edpb.europa.eu/
[7]https://edps.europa.eu/about-edps_en
[8]https://www.torproject.org/

## 5 PRIVACY POLICIES AND IP ADDRESSES

Through privacy policies [2], users are able to access information on the purposes of data processing, on the types of data collected, among other information (Articles 13, 14 GDPR). Therefore, if an organization is collecting/processing IP addresses, they need to mention it in their privacy policies. Accordingly, we have visited and analysed the privacy policies section of each website considered in our study. This analysis of privacy policies and the labelled categories was performed by one legal scholar and one computer scientist. Our findings are summarized in Table 2. From this analysis, we have identified three different ways to handle IP addresses in the websites privacy policy: (i) We observed that IP addresses are processed by the visited websites as these are mentioned explicitly in the website's privacy policy. (ii) Privacy policies can also mention that IP addresses are processed but then they are anonymized. The technique used to anonymize them is never mentioned or detailed. Still, it can be acknowledged that a website considers an IP address as personal data because anonymization consists in the process of turning personal data into data that does not relate to an identified or identifiable person any longer (Recital 26 of the GDPR). (iii) Privacy policies also state that IP addresses are not collected. Optimistically, one can acknowledge that websites consider that IP addresses are personal data. They prefer to mention explicitly that they are not processing IP addresses because they are personal data.

The European Data Protection Board [18] stated that it is crucial to evaluate the *"purpose pursued by the data controller in the data processing"*. Accordingly, we analysed the purposes for processing IP addresses described in the consulted privacy policies. We noticed that the purpose of the collection and processing of IP addresses include the following: *enhancing the user experience and security*. Enhancing the user experience refers to the identification of the location of the user, personalizing and improvement of products, customization of services and trend analysis or the website administration. Some organizations collect IP addresses for security reasons to protect their business against fraudulent behavior, or in case of legal process relating to a criminal investigation or alleged or suspected illegal activity. In fact, these mentioned purposes require some degree of user personalization. As the 29WP refers [18], *'to argue that individuals are not identifiable, where the purpose of processing is precisely to identify them, would be a sheer contradiction in terms. Therefore, the information should be considered as relating to identifiable individuals and the processing should be subject to data protection rules'*. As such, we reason that all these purposes potentially enable the collection of data that conducts to the identification of a user without unnecessary or disproportional effort.

| Description | # Private | # Public | Personal data |
|---|---|---|---|
| Process | 60 | 12 | Yes |
| Anonymize | 1 | 8 | Yes |
| Do not collect | 0 | 3 | Yes |
| Do not mention | 10 | 23 | Unknown |
| No page found | 3 | 4 | Unknown |

Table 2: Analysis of privacy policies of 124 organizations.

*Summary*: We were unable to find the privacy policies of the seven websites. 23 (46%) public websites and 10 (13%) private websites do not mention IP addresses in their privacy policies. It is unclear whether these websites do not store IP address, or whether they do not take care to mention them in their privacy policies. Using our analysis, we found that most private websites (82%) considers that IP addresses are personal data and only 46% of the public websites share this view. Majority of the private websites mention explicitly the processing of IP addresses. Whereas, the situation is slightly different for public websites: 8 of them anonymize IP addresses and 3 of them explicitly state that they do not collect IP addresses.

## 6 IP-BASED SUBJECT ACCESS REQUESTS

We observed that it is possible to conclude that IP addresses are personal data according to the privacy policies of websites (Section 5). The GDPR provides the right to a data subject to access the personal data that a company collects. But do websites accept to answer a request based on an IP address? A subject access right request (SAR) is a request sent by a data subject to a data controller to exercise his/her right to access their data (Article 15 of the GDPR). Several studies [10–13, 35, 41, 44, 48] have used SAR as a methodological tool to assess the transparency of certain data processing, the strength of their authentication procedure or their readiness to comply with the GDPR. During our analysis, we have submitted IP-based subject access requests to private and public organizations.

We have devised a generic subject access request to all the companies and public websites. The full text of the subject access request letter appears in Appendix 9.1. We used this letter to send a SAR to all the organizations. We have complied with all the requests for additional information (like the copy of an ID) to authenticate the SAR made by these websites. Each recipient of our SAR was then permitted one month time to respond to our request as mandated by the GDPR.

We faced several challenges while exercising the SAR for both private companies and public organizations. We visited 74 websites of *private companies* and we submitted SAR to only 62 of them. Four websites (addthis.com, dongao.com, nesine.com and sunstar.com.ph) published wrong e-mail addresses and 6 websites (Google, Youtube, Amazon, LinkedIn, grainger.com and constantcontact.com) did not provide a contact e-mail to submit a SAR if not logged into a user account. Two websites (sprint.com and iheart.com) allow a SAR only for USA citizens. For the *websites of public organizations*, we have visited 50 websites and submitted SAR to only 47 of them. Websites of the Irish and Dutch DPAs did not provide an e-mail address to submit a SAR, while the Italian DPA has provided a wrong e-mail address.

The responses received for our SAR were grouped into 5 different categories. Table 3 shows the categories and the number of organizations related to each of these categories. From 109 organizations to which we submitted a SAR, only 62 thereof have responded (36 private companies and 26 public organizations). Figure 1 in Appendix 9.5 provides the list of private companies and public organizations belonging to each of these categories. In the following sections we depict the obtained responses per category along with its legal analysis.

| Answer's category | Private | Public |
|---|---|---|
| No reply | 26 | 21 |
| No, we have nothing about you | 7 | 23 |
| No account was found | 20 | 0 |
| No, it does not allow to identify you | 4 | 0 |
| No, others | 5 | 3 |

**Table 3: # of websites categorized based on their responses obtained for our IP-based SAR.**

### 6.1 No reply from the website

47 websites (26 private and 21 public) did not reply to the SAR. It is unknown why companies and DPAs did not answer to a SAR request. One could be tempted to conclude that they did not have a process in place to respond to subject access requests. This is particularly surprising (if not shocking) for DPAs. However, the GDPR mandates that the data controllers have an explicit *obligation to facilitate the exercise of data subject rights* (Articles 12(1) and 28(3)(e)), including facilitating SAR requests. Recital 7 recalls that each person should have control of their own personal data and a no-reply to a SAR consists in an obstruction to such control. Recital 59 thereto emphasises that *"modalities should be provided for facilitating the exercise of the data subject's rights"*.

### 6.2 No, we have nothing about you

30 websites (7 private and 23 public) answered they do not have any data matching our request. We compared these answers with the privacy policies of these websites (see Table 4). Some websites (6 private companies and 4 DPAs) are not consistent. Their privacy policies mention the processing of IP addresses, though they fail to find any data relating to the IP addresses included in the SAR request.

| Description | # Private | # Public | Consistency |
|---|---|---|---|
| Process | 6 | 4 | No |
| Anonymize | 0 | 5 | Yes |
| Do not mention | 1 | 12 | Yes |
| Do not collect | 0 | 2 | Yes |

**Table 4: Consistency between the websites privacy policies and their response to our SAR.**

The answers from the remaining websites are consistent with their privacy policies because they either do not mention IP addresses, do not collect IP addresses or anonymize them. In this latter case, a website is indeed unable to recover data corresponding to our IP addresses.

### 6.3 No account was found

20 websites explained that they cannot process the SAR request because they did not find any account corresponding to our request in their system. We expected that websites would use our IP addresses to query their information system and then extract the corresponding information associated to these given IP addresses. But it appears that many of them have queried their information system using the email address used to submit the subject access request. As they have found nothing corresponding to this email address, they reply that they have no data associated to this account or that we need to login or provide our login information to complete our request. For instance, *Roblox* replied to us that: *You must verify ownership of any associated Roblox account. We must have a Roblox user name in order to proceed with a GDPR request.*

This kind of answer shows how websites process subject access requests. They have completely ignored the IP addresses provided in our requests to focus only on the email address used to send the request. Thus, their procedure was not able to treat our request.

It looks like, their implementation is (maybe) motivated by the need to authenticate the request: *The controller should use all reasonable measures to verify the identity of a data subject who requests access, in particular in the context of online services and online identifiers* (Recital 64 of the GDPR). They have email addresses in their information systems and they want to ensure that they provide data only to a legitimate user.

Internet users need to provide many personal data, like their last name, first name, gender, location, phone number, email addresses etc.. to create an account on a website. Even if it is possible to lie for some of these fields, it is tempting to visit websites without creating an account: it is faster and also a privacy-preserving choice. However, it implies that the websites can do whatever they want with the user's data as there is no way for a user to exercise the rights provided by the GDPR. This kind of answer is therefore problematic because it creates a legal loophole.

### 6.4 No, it does not allow to identify you

Four websites (Tripdvisor, Microsoft, lyst.co.uk and rubiconproject.com) have taken into account the IP addresses provided in our request. However, according to them, they cannot process the request, since IP addresses are shared and dynamic. These websites have considered that an IP address can be used by multiple users at the same time and thus they are unable to distinguish, in their own information system, the data belonging to us from the data of other users using the same IP addresses. In such situation, if a website is able to demonstrate that it is not in a position to identify a concrete user, it can deny the request (Article 11(2)).

Additionally, there is a *risk* for the websites to disclose the information of other users and whenever such a personal data breach risk exists (Article 4 (12)), a website can deny the request. The UK DPA alerts that *the level of checks a data controller should make may depend on the possible harm and distress that inappropriate disclosure of the information could cause to the individual concerned* [9]. Recital 63 of the GDPR adverts that such *" right should not adversely affect the rights or freedoms of others"*

One may argue that these websites could have requested *more identifiable information* from the user in order to identify the requester (as mentioned in section 3.2).

In effect, Recital 64 of the GDPR states that the controller should use all reasonable measures to verify the identity of a data subject who requests access, in particular in the context of online services and online identifiers. This vague concept of "reasonable measures" might result in data controllers implementing weak or irrelevant identity verification means upon receiving a request.

To note, a website is not obliged to collect additional information to identify the data subject for the sole purpose of complying with the GDPR subject access rights (Article 11 (2) and Recital 57 of the GDPR). This argument excludes this possibility of acquiring additional data when it is no longer necessary.

### 6.5 No with a data breach

The answer for rubiconproject.com is a bit different from the answers from Tripdvisor, Microsoft, lyst.co.uk. *rubiconproject.com* acknowledged the presence of our IP addresses in its log file. The reply states: *we do process data associated with IP addresses XX.XX.XX.XX[9] and YY.YY.YY.YY[9], our searches suggest that these addresses are associated with multiple different devices across multiple territories. This indicates that these IP addresses are used by multiple different users etc.* The response shows that there is a particular process followed by this website to handle IP-based SAR request. Even if they only acknowledge the presence of our IP addresses, it is already a data breach (Article 4(12) GDPR). We could have put any arbitrary IP addresses in our request to learn if the addresses are present in the log file of rubiconproject.com. If other websites (like Alcoholics Anonymous www.aa.org) use the same procedure to reply to our request, sensitive information of Internet users could be exposed to anybody submitting an IP-based subject access request.

### 6.6 No, Others

8 websites (5 private and 3 public) did not provide any useful information in their response to the SAR. Generally, these websites requested additional data to identify the requester.

This implies that IP addresses alone were not a sufficient piece of data to identify the requester.

Such request for added data is in line with the GDPR wherein a controller – having reasonable doubts concerning the identity of a person making the request – may request the provision of additional information necessary to confirm the identity of the data subject (Article 12(6)).

However, there seems to be no substantive reason for a data subject to reveal the real identity to these websites through the requested documents: original signed letter (Luxembourgian DPA), SAR in the official language of the DPA (Austrian DPA), or a INE number, *i.e.*, a statistical number (orange.es). These documents requested by these DPAs must be *proportional* or *necessary* to the website's knowledge of the data subject. Moreover, the *minimization principle* mandates that personal data shall be adequate, relevant and limited to what is necessary in relation to the purposes for which they are processed (Article 5 (1) (c)). Recital 39 specifies further that *personal data should be processed only if the purpose of the processing could not reasonably be fulfilled by other means.* The "necessity" or "proportionality" requirement that both these provisions note, refers to both quantity and also to the quality of personal data. It is then clear that these DPAs should not process excessive data if this entails a disproportionate interference in the data subject's rights, and hence, a privacy invasion.

---

[9]Anonymized

*Summary*: In Table 5, we can contrast the consistency between what companies and organizations publish on their privacy policies and the way they respond to the IP-based SAR. There is only one private company and just 17 public organizations that are consistent. After analysing all the responses received by websites, we can see that IP-based SAR requests are not handled properly.

In the next section 7 we propose some solutions that could be implemented in-order to properly exercise IP-based SAR.

## 7 HOW TO FIX IP-BASED SARS?

The situation of IP-based subject access requests is similar to cookie-based subject access requests. In both the studies on cookie-based subject access requests [11, 48], it was highlighted that it was not possible for a user to claim that he/she has used a given tracking cookie while visiting the websites. Despite this critical issue, several advertising agencies replied positively to researchers requests. This behavior triggered data breaches. Still, it helped to understand how online tracking works and how the GDPR impacts online tracking. There is clearly a conflict between security (risk of data breaches) and transparency.

A first solution would be to change how IP addresses are allocated. Any Internet user should be able to create his/her own IP address such that she can later prove that he/she has used this address to a data controller. The computation of an IP address can be based on the user's public key for instance. Then, the IP address is computed as a cryptographic commitment [16]. When the user wants to access his/her data, then he/she provides the cryptographic materials to let the data controller open the commitment. The data controller can then determine if the elements provided by the user help to verify that the request of the user is legitimate. Such a stateless IP address allocation scheme has been proposed in the past for IPV6 addresses [39, 40]. These schemes[39, 40] are security oriented but it appears that they are useful to create GDPR compliant identifiers. It is not realistic to believe that we can change how IP addresses are allocated, as it will create plenty of routing issues. However, it is important for people defining future networking technologies to take into consideration legal issues and that on-the-shelf solution exists like [39, 40].

Another solution to fix IP-based subject access request consists to ask the assistance of data protection authorities and Internet Service Providers. In this case, this kind of solution will work for researchers to expose how websites collect and process data from accountless users. As a first step, a researcher can ask a data protection authority to certify that the researcher is the only user associated to a given IP address. Then, the researcher can contact his/her ISP to provide a certificate of usage. This certificate will be signed using the ISP secret key and verified by the data controller using the ISP public key. The data protection authority role is to establish the link between the IP address and the researcher. The ISP is used as a certification authority which recognises the authority of the data protection authority and which is recognized by the data controllers (*i.e.* websites) accordingly. A similar approach has been used in[48] for cookie with affidavit and in VICEROY [29] which is a similar approach based on SGX technology. The certificate would attest that a given individual has used a certain IP address for a given period of time. Such a certification scheme would need to rely on public key infrastructure to attest that it was created by an authorized ISP. We have attempted on our own to ask an ISP to provide us such a certificate during our study. Unfortunately, our request was never answered.

## 8 CONCLUSION

Our survey of 109 websites show that websites consider IP addresses as personal data in their privacy policies. However, it is not possible to access any data related to the IP addresses used while visiting their websites as an accountless user. This is due to the current network topology and the way IP addresses are allocated. Currently, *IP addresses as personal data* is only a theoretical statement for Internet users because in practice there is no practical means for them to exercise their rights. Internet users cannot prove that they have used an IP address. Therefore, it is easy for websites to deny IP-based subject access requests.

Fixing IP-based subject access requests is difficult because the risk of a data breach cannot be omitted. It is however important to fix it in order to achieve transparency for accountless Internet users and to understand the extend of online tracking. We hope that our work can help shape the design of future networking technologies in order to make them privacy compliant.

|  | # of websites (private) | No reply | No, We have nothing about you | No account was found | No, it does not allow to identify you | No, others | # of websites (public) | No reply | No, We have nothing about you | No account was found | No, it does not allow to identify you | No, others |
|---|---|---|---|---|---|---|---|---|---|---|---|---|
| **Process** | 60 | 19 | 6 | 17 | 3 | 5 | 12 | 7 | 6 | 0 | 0 | 0 |
| **Anonymize** | 1 | 1 | 0 | 0 | 0 | 0 | 8 | 3 | 4 | 0 | 0 | 0 |
| **Do not collect** | 0 | 0 | 0 | 0 | 0 | 0 | 3 | 0 | 3 | 0 | 0 | 0 |
| **Do not mention** | 10 | 4 | 1 | 3 | 1 | 0 | 23 | 8 | 10 | 0 | 0 | 3 |
| **No page found** | 3 | 2 | 0 | 0 | 0 | 0 | 4 | 3 | 0 | 0 | 0 | 0 |
| **Total** | 74 | 26 | 7 | 20 | 4 | 5 | 50 | 21 | 23 | 0 | 0 | 3 |
|  | **Private companies** | | | | | | **Public Organizations** | | | | | |

Table 5: Consistency between the websites privacy policies and their response to our SAR. (SAR sent 62- Private companies and 47 public organizations)

# 9 APPENDIX
## 9.1 SAR Template used in our experiments

We have created a small template with relevant information for the Data controllers. Along with first and last name, we provide a set of IP addresses used to access their website using different networks.

Dear Data Controller,

I am hereby requesting a copy of all my personal data held and/or undergoing processing, according to Article 15 of the GDPR. Please confirm whether or not you are processing personal data concerning me. In case you are, I am hereby requesting access to the following information: All personal data concerning me that you have stored. This includes any data derived about me, such as opinions, inferences, settings and preferences.

Please make the personal data concerning me, which I have provided to you, available to me in a structured, commonly used and machine-readable format, accompanied with an intelligible description of all variables.

I am including the following information necessary to identify me:

Name - first-name last-name
IP addresses used to access are as follows-
XX.XX.XX.XXX
XXXX:XXXX:XXXX:XXXX::XXX:XX

Yours sincerely,

first-name

(As laid down in Article 12(3) GDPR, you have to provide the requested information to me without undue delay and in any event within one month of receipt of the request. According to Article 15(3) GDPR, you have to answer this request without cost to me.)

### 9.2 List of private companies

We have chosen 22 popularly used websites across the globe and we have chosen 52 websites of companies that set cookies on their user's browser. The computation of these cookies depends on the IP address of the user. Table 6 shows a list of these companies.

### 9.3 List of public organizations

We have considered 50 public organizations that includes 48 DPAs, EDPB and EDPS. It is interesting to see how they process IP-addresses and how they respond to a IP-based SAR. Table 7 provides the name of all the public organizations.

### 9.4 Privacy policies- IP address is personal data or not

We have visited the privacy policies of 124 organizations to check if these organizations consider IP address personal data or not. Table 8 and 9 provides list of private companies and public organizations based on different categories. Each of these categories shows if IP addresses are treated as personal data or not.

### 9.5 Summary of responses to SAR

Figure 1 provides the list of different private companies and public organizations and the types of responses we received for our IP-based SAR.

| Popular companies | | | | | |
|---|---|---|---|---|---|
| Google | Youtube | Amazon | LinkedIn | Reddit | Indeed |
| Zoom | Yahoo | Ebay | Pinterest | Wikipedia | Twitter |
| Twitch | Roblox | Bitly | Fandom | Tripadvisor | Apple |
| Microsoft | Netflix | Euronews | Facebook | | |
| Companies that set cookies using user's IP address | | | | | |
| 1000.menu | adbox.lv | addthis.com | admanmedia.com | adswizz.com | assets.new.siemens.com |
| bigcommerce | britishairways.com | caranddriver.com | dongao.com | constantcontact.com | duda.co |
| dsar.everydayhealth.com | forever21.com | gismeteo.ua | grainger.com | gumgum.com | iheart.com |
| jpnn.com | kuleuven.be | louisvuitton.com | lifepointspanel.com | lyst.co.uk | mckinsey.com |
| nesine.com | mylu.liberty.edu | my-personaltrainer.it | okta.com | officedepot.com | orange.es |
| pgatour.com | pubmatic.com | point2homes.com | russianfood.com | rubiconproject.com | sinoptik.ua |
| spiceworks.com | smartadserver.com | sprint.com | start.me | sunstar.com.ph | trafficjunky.net |
| turktelekom.com.tr | urbanfonts.com | vans.com | warriorplus.com | wikimedia.org | wikiquote.org |
| worldpopulationreview.com | yandex.com.tr | yandex.kz | zoho.com | | |

**Table 6: List of 74 private companies visited as external user (22 popular companies and 52 companies that set cookies using IP address).**

| Public organizations | | | |
|---|---|---|---|
| ADA (Lithuania) | AEPD (Spain) | AKI (Estonia) | BfDI (Germany) |
| ANSPDCP (Romania) | AP (Netherlands) | APD/GBA (Belgium) | CNPD (Luxembourg) |
| BayLfD (Bavaria, public sector) | AZOP (Croatia) | BayLDA (Bavaria) | HDPA (Greece) |
| UOOU (Czech Republic) | BlnBDI (Berlin) | CNIL (France) | TLfDI (Thuringia) |
| LfDI (Mecklenburg-Vorpommern) | CNPD (Portugal) | Commissioner (Cyprus) | ULD (Schleswig-Holstein) |
| LfDI (Baden-Württemberg) | Datainspektionen (Sweden) | Datatilsynet (Denmark) | NAIH (Hungary) |
| Datatilsynet (Norway) | LDI (North Rhine-Westphalia) | UOOU (Slovakia) | Persónuvernd (Iceland) |
| Datenschutzzentrum (Saarland) | DPC (Ireland) | DSB (Austria) | DSB (Saxony) |
| LFDI (Rhineland-Palatinate) | DVI (Latvia) | LfD (Lower Saxony) | UODO (Poland) |
| Datenschutzstelle (Liechtenstein) | IDPC (Malta) | HBDI (Hesse) | CPDP (Bulgaria) |
| Tietosuojavaltuutetun toimisto (Finland) | HmbBfDI (Hamburg) | ICO (UK) | LfDI (Bremen) |
| Garante per la protezione... (Italy) | IP (Slovenia) | LDA (Brandenburg) | LfD (Saxony-Anhalt) |
| EDPB | EDPS | | |

**Table 7: List of 50 public organizations (48 DPAs and EDPB, EDPS).**

| Process IP address | | | | | |
|---|---|---|---|---|---|
| Google | Youtube | Amazon | LinkedIn | Reddit | Apple |
| Zoom | Yahoo | Indeed | Pinterest | Wikipedia | Twitter |
| Twitch | Roblox | Bitly | Fandom | Microsoft | Euronews |
| Netflix | Facebook | Instagram | worldpopulationreview.com | caranddriver.com | officedepot.com |
| yandex.kz | jpnn.com | turktelekom.com.tr | grainger.com | vans.com | britishairways.com |
| yandex.com.tr | orange.es | gumgum.com | lyst.co.uk | duda.co | trafficjunky.net |
| sinoptik.ua | zoho.com | constantcontact.com | sprint.com | okta.com | point2homes.com |
| addthis.com | pubmatic.com | wikiquote.org | iheart.com | mckinsey.com | rubiconproject.com |
| start.me | my-personaltrainer.it | nesine.com | pgatour.com | smartadserver.com | dsar.everydayhealth.com |
| spiceworks.com | forever21.com | adswizz.com | wikimedia.org | bigcommerce | lifepointspanel.com |
| admanmedia.com | | | | | |
| **Anonymize IP address before processing** | | | | | |
| gismeteo.ua | | | | | |
| **Do not mention IP address in the privacy policy** | | | | | |
| louisvuitton.com | mylu.liberty.edu | sunstar.com.ph | kuleuven.be | warriorplus.com | assets.new.siemens.com |
| urbanfonts.com | 1000.menu | Ebay | Tripadvisor | | |
| **Privacy policy page not found** | | | | | |
| adbox.lv | russianfood.com | dongao.com | | | |

Table 8: 74 Companies categorised based on their privacy policies.

| Process IP address | | | | | |
|---|---|---|---|---|---|
| Slovakia | Poland | Schleswig-Holstein | Iceland | Brandenburg | North Rhine-Westphalia |
| Lower Saxony | Saxony-Anhalt | Bremen | Rhineland-Palatinate | Slovenia | Hesse |
| **Anonymize IP address before processing** | | | | | |
| Germany | Liechtenstein | Bavaria | Lower Saxony | Bremen(LFDI) | LFD (Saxony-Anhalt) |
| EDPB | EDPS | | | | |
| **Do not collect/process IP address** | | | | | |
| Schleswig-Holstein | Saarland | Brandenburg | | | |
| **Do not mention IP address in the privacy policy** | | | | | |
| Estonia | Belgium | Croatia | Cyprus | France | Luxembourg |
| Sweden | Ireland | Austria | Saxony | Latvia | Denmark |
| Norway | Greece | Hamburg | UK | Malta | Mecklenburg-Vorpommern |
| Baden-Württemberg | Hungary | Finland | Thuringia | Czech Republic | |
| **Privacy policy page not found** | | | | | |
| Lithuania | Spain | Romania | Netherlands | | |

Table 9: 48 DPAs, EDPB, EDPS categorised based on their privacy policies.

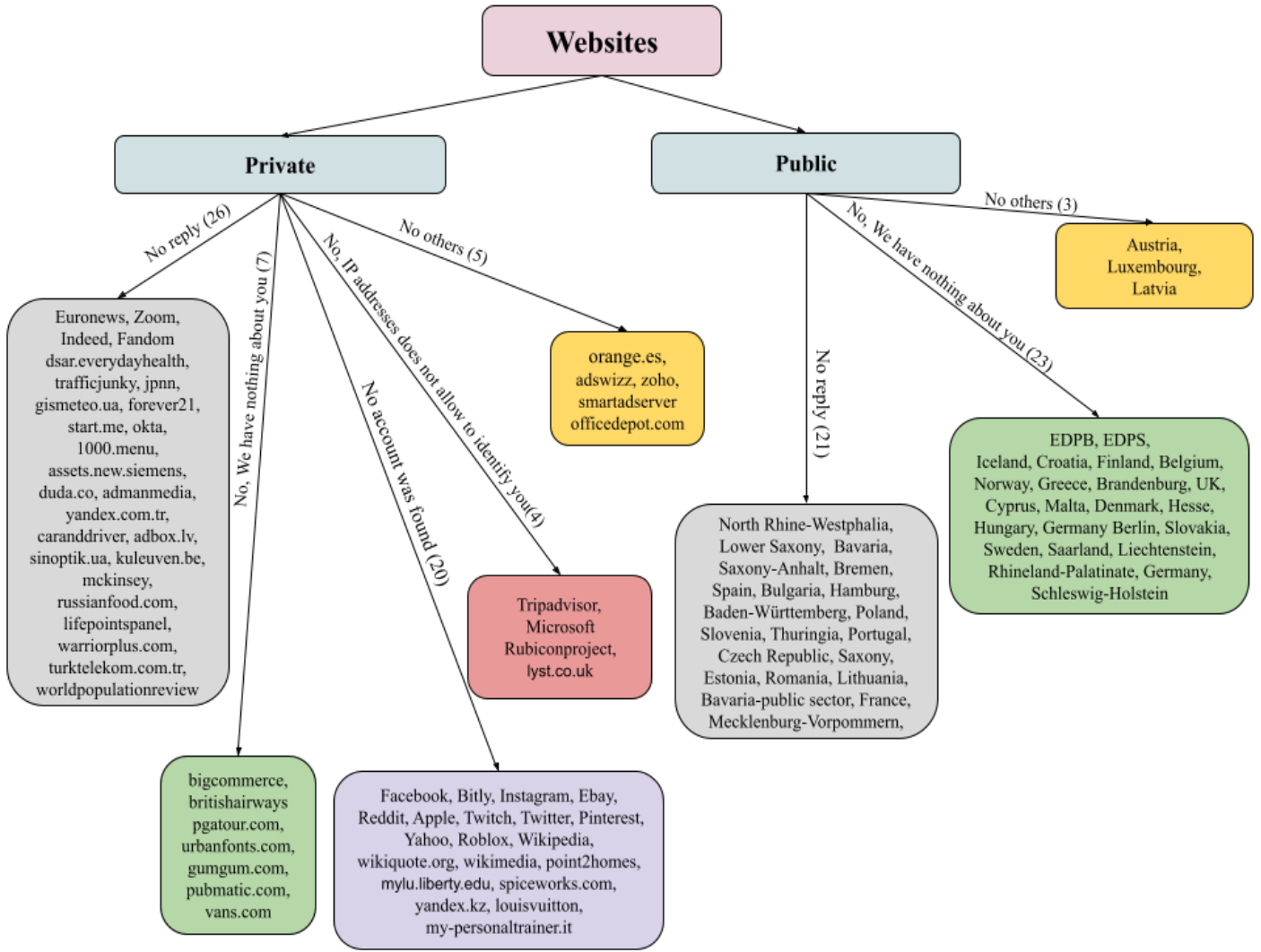

Figure 1: Responses obtained by websites of private and public organizations for a IP-based SAR request.